\documentstyle[preprint,epsfig,aps]{revtex}

\begin{document}

\title{Consequences of a $\Lambda_{c}/D$ enhancement effect on the non-photonic electron nuclear modification factor in central heavy ion collisions at RHIC energy}

\date{\today}

\author{G.~Mart\'{\i}nez-Garc\'{\i}a$^1$\thanks{Gines.Martinez@subatech.in2p3.fr}, S.~Gadrat$^1$\thanks{Sebastien.Gadrat@subatech.in2p3.fr} and 
P.~Crochet$^2$\thanks{Philippe.Crochet@clermont.in2p3.fr}}

\address{
$^{1}$SUBATECH
(IN2P3/CNRS - Ecole des Mines - Universit\'e de Nantes) Nantes, France\\
$^{2}$Laboratoire de Physique Corpusculaire 
(IN2P3/CNRS - Universit\'{e} Blaise Pascal) Clermont-Ferrand, France}
\maketitle

\begin{abstract}
The RHIC experiments have measured the nuclear modification factor 
$R_{AA}$ of non-photonic electrons in Au+Au collisions at 
$\sqrt{s_{NN}} = 200~{\rm GeV}$. This $R_{AA}$ exhibits a large suppression 
for $p_{\rm t} > 2~{\rm GeV}/c$ which is commonly attributed to 
heavy-quark energy loss. 
It is expected that the heavy-quark radiative energy loss 
is smaller than the light quark one because of the so-called dead-cone effect.
An enhancement of the charm baryon yield with respect to the charm meson yield,
as it is observed for light and strange hadrons, can explain part of the 
suppression. 
This phenomenon has been put forward in a previous work.
We present in this paper a more complete study based on a detailed 
simulation which includes electrons from charm and bottom decay, charm and 
bottom quark realistic energy loss as well as a more realistic modeling 
of the $\Lambda_c/D$ enhancement.
We show that a $\Lambda_{c}/D$ ratio close to unity, as observed 
for light and strange quarks, could explain $20-25\%$ of the suppression of 
non-photonic electrons in central Au+Au collisions. 
This effect remains significant at relatively high non-photonic electron 
transverse momenta of $8-9~{\rm GeV}/c$.
\end{abstract}

\pacs{12.38.Mh, 25.75.-q, 25.75.Nq}
\keywords{Heavy-ion collisions, nuclear modification factor, 
heavy-quark energy loss, baryon-to-meson ratio, non-photonic electrons}

\newpage
One of the most robust experimental evidence for the creation of a new 
state of matter in heavy ion collisions at the Relativistic Heavy Ion 
Collider (RHIC) is the large suppression of light hadrons at 
high transverse momentum ($p_{\rm t}$)~\cite{RAA_compil}.
This phenomenon is well reproduced by models 
which take into account the radiative energy loss of high $p_{\rm t}$ light 
quarks and gluons propagating through a dense medium of colored quarks and 
gluons~\cite{Salgado:2005pr}. Further insights into the underlying mechanism 
can be obtained from the study of heavy hadrons.
In contrast to intermediate-$p_{\rm t}$ light hadrons which are predominantly 
produced by gluon fragmentation, charm and bottom hadrons originate 
from the fragmentation of heavy quarks.
Quarks are supposed to lose less energy than gluons in the medium due to a 
smaller color charge coupling.
In addition, radiative energy loss was predicted to be smaller 
for heavy quarks as compared to light quarks because of the so-called 
``dead-cone'' effect which limits the medium induced radiative energy loss
at forward angles~\cite{Dokshitzer:2001zm}.
Surprisingly, recent data from the \textsc{PHENIX} and the \textsc{STAR} 
collaborations in Au+Au collisions at $\sqrt{s_{NN}} = 200~{\rm GeV}$ show 
that the quenching of heavy quarks, as studied indirectly via the so-called 
non-photonic electrons\footnote{In contrast to light hadrons, 
the heavy flavor quenching is, so far, not measured experimentally through 
identified hadrons, but in an inclusive way via the nuclear modification 
factor ($R_{AA}$) of non-photonic electrons. 
The latter is obtained from the $p_{\rm t}$ distributions of electrons 
(after subtraction of Dalitz-decay electrons from light hadrons and 
photon-conversion electrons) in $AA$ collisions 
(${\rm d}N^e_{AA}/{\rm d}p_{\rm t}$) and in $pp$ collisions
(${\rm d}N^e_{pp}/{\rm d}p_{\rm t}$) as:
$$R_{\rm AA} = \frac{{\rm d}N^e_{AA}/{\rm d}p_{\rm t}}{<N^{AA}_{\rm coll}>{\rm d}N^e_{pp}/{\rm d}p_{\rm t}}$$
where $<N^{AA}_{\rm coll}>$ is the average number of nucleon-nucleon
collisions corresponding to a given centrality class.},
is stronger than theoretical 
expectations~\cite{Adler:2005xv,Adare:2006nq,Abelev:2006db}
and is as large as that of light mesons.
Reconciling these data with model predictions is a real challenge which 
triggers a lot of theoretical activities nowadays.
Only models which assume a very large medium opacity~\cite{Armesto:2005mz}, 
an additional collisional energy loss~\cite{Djordjevic:2006kw}
or effective energy loss from multiple fragmentations and dissociations of 
heavy quarks and mesons ($D$ and $B$) in the medium~\cite{Adil:2006ra} 
can describe, with a relatively good agreement, the data (for a recent review, 
see~\cite{Ralf}).

In this paper we investigate the possibility that part of 
the strong suppression of non-photonic electrons might be due to another
source of electrons, namely charmed baryons.
Indeed, whereas light mesons are largely suppressed in heavy ion 
collisions at RHIC, the suppression of non-strange and strange baryons
is observed to be much less in the intermediate $p_{\rm t}$ range
($2 < p_{\rm t} < 4~{\rm GeV}/c$)~\cite{RBM_compil}.
This is commonly referred to as the anomalous baryon/meson enhancement.
This anomalous baryon/meson enhancement is relatively well understood in the 
framework of 
the recombination model which assumes that, at low and intermediate 
$p_{\rm t}$, hadronization occurs via the coalescence of ``free" quarks 
(and anti-quarks)~\cite{coalescence_compil}.
An anomalous baryon/meson enhancement for charm
hadrons leads naturally to a non-photonic electron $R_{AA}$ smaller than
one.
This is mostly due to a smaller semi-leptonic decay branching ratio of 
charm baryons ($\Lambda_{c}$) as compared to charm 
mesons (see Tab.~\ref{tab_PDG}). As a consequence, part of the 
experimentally measured $R_{AA}$ of non-photonic electrons should not be 
attributed to energy loss. 
We show that the $\Lambda_c/D$ enhancement can explain up to 25\% 
of the non-photonic electron suppression data measured by the 
\textsc{PHENIX} collaboration in Au+Au collisions at 
$\sqrt{s_{NN}} = 200~{\rm GeV}$~\cite{NewPhenix}.

The main assumption we put forward is that, in a deconfined medium, charm
baryon production is enhanced relative to charm meson production, as 
compared to the vacuum.
This assumption is qualitatively justified in the framework of the 
recombination model.
Although this model does not provide detailed predictions on charm hadron 
production yet, it 
successfully describes the (non-charm) baryon/meson enhancement measured in 
Au+Au collisions at $\sqrt{s_{NN}} = 200~{\rm GeV}$.
A relatively good agreement is obtained not only for the light hadron ratio 
$p/\pi^+$, but also for heavier hadron ratios such as $\Lambda/K^0_s$ and 
$\Omega/\phi$~\cite{Sarah}. 
Extrapolating these results to charm hadrons is not 
straightforward because the mass of the charm quark is much larger than 
that of light and strange quarks. 
The consequences are threefold as far as the recombination mechanism is 
concerned. 
First, whereas the $p_{\rm t}$ of a light baryon (meson) amounts to 
3 (2) times the initial $p_{\rm t}$ of its valence quarks, the $p_{\rm t}$ of 
a (single) charm baryon or a charm meson is likely to be very close to that of 
the charm quark. 
Secondly, considering a light quark and a heavy quark with the same velocity 
(which is the essential requirement for the coalescence 
process to take place~\cite{Lin:2003jy}), 
the heavy quark momentum is much larger than that of light partons.
As a consequence, one can expect the enhancement of the 
charm baryon/meson ratio to appear at higher $p_{\rm t}$
than that of non-charm hadrons. 
The recombination model indeed predicts, for non-strange and strange
hadrons, that the heavier the hadron, the larger the $p_{\rm t}$ of the 
baryon/meson enhancement.
This has been observed for non-strange and strange baryon/meson ratios
by the \textsc{STAR} collaboration~\cite{Sarah}.
Finally, the fragmentation time of heavy quarks is small as compared
to light quarks. 
According to~\cite{Adil:2006ra}, the formation time of 
a $10~{\rm GeV}/c$ pion, $D$ meson and $B$ meson is $20$, $1.5$ and 
$0.4~{\rm fm}/c$, respectively and it is as small as 
$\sim 3~{\rm fm}/c$ for a $\Lambda_c$ with $p_{\rm t} = 20-30~{\rm GeV}/c$.
Due to these considerations, it is obvious that the baryon/meson enhancement
for non-charm hadrons and charm hadrons can be significantly different.
In the following, we only assume that, in view of experimental 
results on the baryon/meson enhancement for non-strange and strange hadrons, 
a similar enhancement is a priori conceivable for charm 
hadrons\footnote{A very recent theoretical study in 
the framework of the recombination model confirms this 
assumption~\cite{Lee:2007wr}.}.
Remarkably, such an enhancement has strong implications on the 
nuclear modification factor of non-photonic electrons.
It leads to a decrease of the yield of non-photonic electrons in 
$A+A$ collisions because, as shown in Tab.~\ref{tab_PDG}, the inclusive 
semi-leptonic decay branching ratio of charm baryons is smaller than that of 
charm mesons.
Therefore, the nuclear modification factor of non-photonic electrons should 
decrease as well.
This can be easily illustrated in the following way.
Assuming that charm production scales with the number of binary collisions
(i.e. $R_{AA} = 1$ in absence of medium effects) and that the relative yields 
of $D$ mesons are the same in $pp$ and in $A+A$ collisions, 
a $p_{\rm t}$ integrated $R_{AA}$ can be calculated for different 
${\cal C}$ enhancement factors, 
${\cal C} = \left( N_{\Lambda_c,\bar{\Lambda}_c}/N_D\right)_{AA} / \left( N_{\Lambda_c,\bar{\Lambda}_c}/N_D\right)_{pp}$ \\
with
\begin{equation}
N_{\Lambda_c,\bar{\Lambda}_c} / N_D = \\
\frac{N_{\Lambda_c} + N_{\bar{\Lambda}_c}}{N_{D^+} + N_{D^-} + N_{D^0} + N_{\bar{D}^0} + N_{D^+_s} + N_{D^-_s}}
\end{equation}
according to 
\begin{equation}
R_{AA} = \\	
\frac{1 + (N_{\Lambda_c,\bar{\Lambda}_c}/N_D)_{pp} }
{1 + {\cal C} (N_{\Lambda_c,\bar{\Lambda}_c}/N_D)_{pp}} \times
	\frac{1 + {\cal C} (N_{e \leftarrow \Lambda_c} / N_{e \leftarrow D})_{pp}}
	{1 + (N_{e \leftarrow \Lambda_c} / N_{e \leftarrow D})_{pp}}
\label{eq1}
\end{equation}
where 
\begin{equation}
N_{e \leftarrow \Lambda_c} / N_{e \leftarrow D} =  
      \frac{(N_{\Lambda_c,\bar{\Lambda_c}}/N_D)BR_{\Lambda_c,\bar{\Lambda_c}}}
           {(N_{D^\pm}/N_D)BR_{D^\pm}+(N_{D^0,\bar{D^0}}/N_D)BR_{D^0,\bar{D^0}}+(N_{D_s^\pm}/N_D)BR_{D^\pm_s}}.
\label{eq2}
\end{equation}
$N$ is the charm hadron yield and $BR$ is the hadron semi-leptonic decay 
branching ratio.
According to Tab.~\ref{tab_PDG}, $N_{\Lambda_c,\bar{\Lambda}_{c}}/N_D=7.3\%$, 
$N_{D^\pm}/N_D=21\%$, $N_{D^0,\bar{D}^0}/N_D=67\%$, and $N_{D_s^\pm}/N_D=12\%$
such that $N_{e \leftarrow \Lambda_c} / N_{e \leftarrow D} = 3.63\%$.
Therefore, an enhancement factor ${\cal C}$ of 12
leads to a non-photonic electron $R_{AA}$ of $0.79\pm0.07$ and 
in the extreme case of an infinite enhancement, the non-photonic electron 
$R_{AA}$ reaches $0.51$.

The above idea has already been proposed in~\cite{Sorensen:2005sm}.
Before going to our simulation results, we present in Tab.~\ref{tab_diff}
the main differences between our approach and the one of 
ref.~\cite{Sorensen:2005sm}.
The choice of a Gaussian shape for the $p_{\rm t}$ dependence of the 
$\Lambda_c/D$ ratio in Au+Au collisions is motivated by results from
the coalescence model for heavy quarks~\cite{Greco:INFN2005}.
For $pp$ collisions, we use the predictions from \textsc{PYTHIA} and not the 
shape from the measured $\Lambda/K^0_{s}$ ratio since experimental results
from the \textsc{STAR} collaboration indicate a strong mass dependence 
of baryon/meson ratios~\cite{Sarah}.
These assumptions lead to a significant difference in the maximum the 
$\Lambda_c/D$ ratio and in its location in $p_{\rm t}$, as it is 
reported later.
Finally, the present work includes a more realistic treatment
of the heavy-quark energy loss as well as the contribution of electrons 
from bottom decay which is ignored in~\cite{Sorensen:2005sm}.

\begin{figure}[htb]
 \centering\epsfig{file=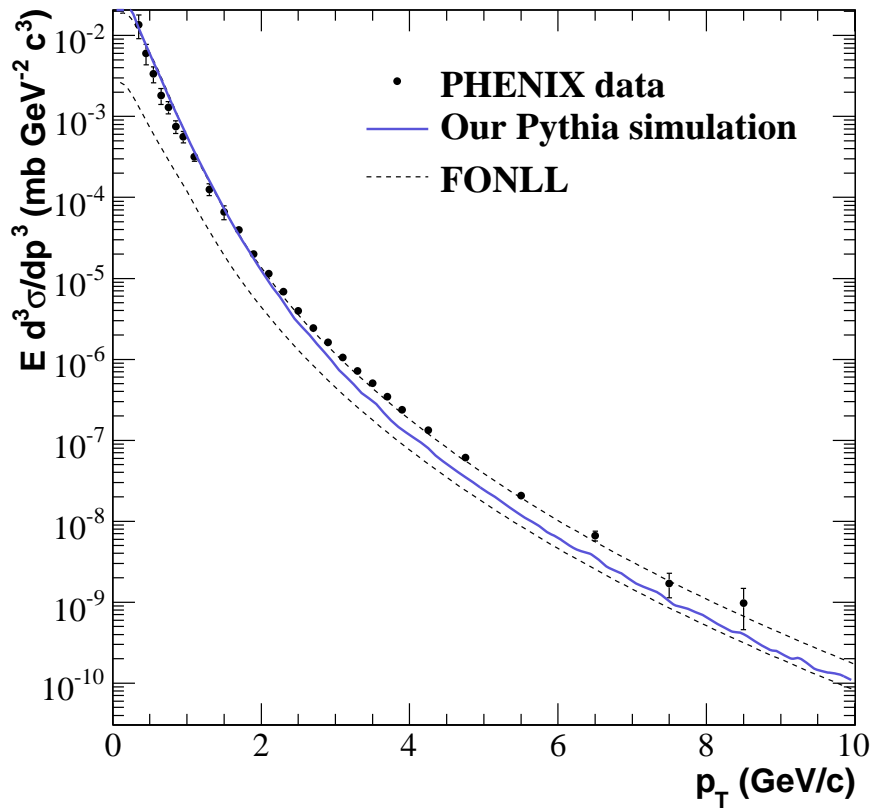,width=0.75\linewidth}
  \caption{Invariant differential cross-section of non-photonic electrons 
	(dots) measured in $pp$ collisions at 
         $\sqrt{s} = 200~{\rm GeV}$~\protect\cite{NewPhenix}.
	 The dashed curves show the prediction from \textsc{FONLL}
	calculations~\protect\cite{Cacciari:2005rk}. 
	The solid curve shows the result of the \textsc{PYTHIA} 
	simulation as described in the text. 
        The simulated spectrum is normalized from the integration of the 
	measured spectrum in the range 
	$1.4 < p_{\rm t} < 4~{\rm GeV}/c$.}
  \label{figure1} 
\end{figure}

Our simulation framework is based on the \textsc{PYTHIA}-6.152 event 
generator~\cite{Sjostrand:1993yb}.
The \textsc{PYTHIA} input parameters were first tuned according 
to~\cite{Adcox:2002cg} and the \textsc{PHENIX} acceptance cut ($|\eta| < 0.35$)
was applied in order to correctly reproduce the $p_{\rm t}$ distribution of 
non-photonic electrons measured in $pp$
collisions at $\sqrt{s} = 200~{\rm GeV}$~\cite{NewPhenix}. 
As it can be seen in Fig.~\ref{figure1}, the agreement between the 
simulation and the data is rather good except in the high $p_{\rm t}$ region 
where the simulation under-predicts the data. 
The result of the simulation is also compared to \textsc{FONLL} 
(Fixed Order Next to Leading Log) predictions~\cite{Cacciari:2005rk}. 
As already observed in~\cite{NewPhenix}, the \textsc{PHENIX} data is 
in agreement with \textsc{FONLL} within the theoretical uncertainties.

\begin{figure}[htb]
  \centering\epsfig{file=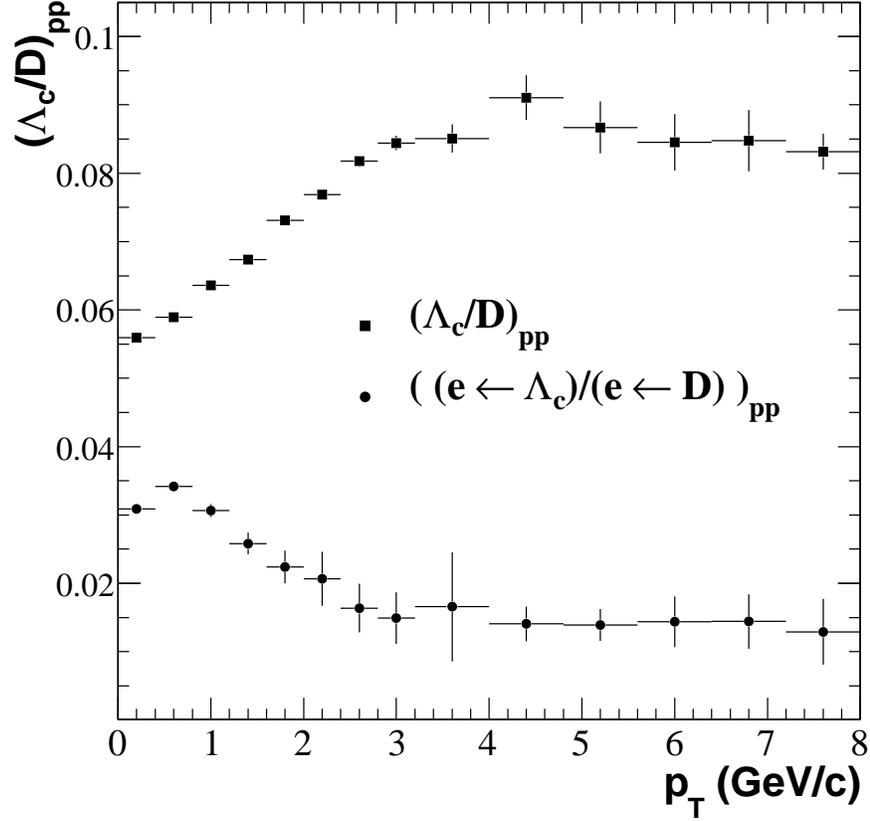,width=0.75\linewidth}
  \caption{Transverse momentum dependence of the charm baryon/meson ratio 
	(squares) and decay electrons from charm baryons over decay electrons
	from charm mesons (dots).
	The results are obtained from the \textsc{PYTHIA} simulation 
	described in the text for $pp$ collisions at 
	$\sqrt{s} = 200~{\rm GeV}$.}
  \label{figure2} 
\end{figure}

Table~\ref{tab_PDG} shows that the $\Lambda_c/D$ ratio 
amounts to 7.3\% (in $4\pi$) which translates to 3.63\% after 
convolution of the species yields with their corresponding semi-leptonic 
decay branching ratio.
On the other hand, Fig.~\ref{figure2} shows that, in the 
$p_{\rm t} > 2~{\rm GeV}/c$ region of interest discussed hereafter, 
this ratio is even smaller 
($\sim 1.5\%$) because the decay electron spectrum of 
$\Lambda_c$ is softer than that of $D$ mesons.
This leads to an additional suppression of the non-photonic electron yield 
at intermediate $p_{\rm t}$.

\begin{figure}[htb]
  \centering\epsfig{file=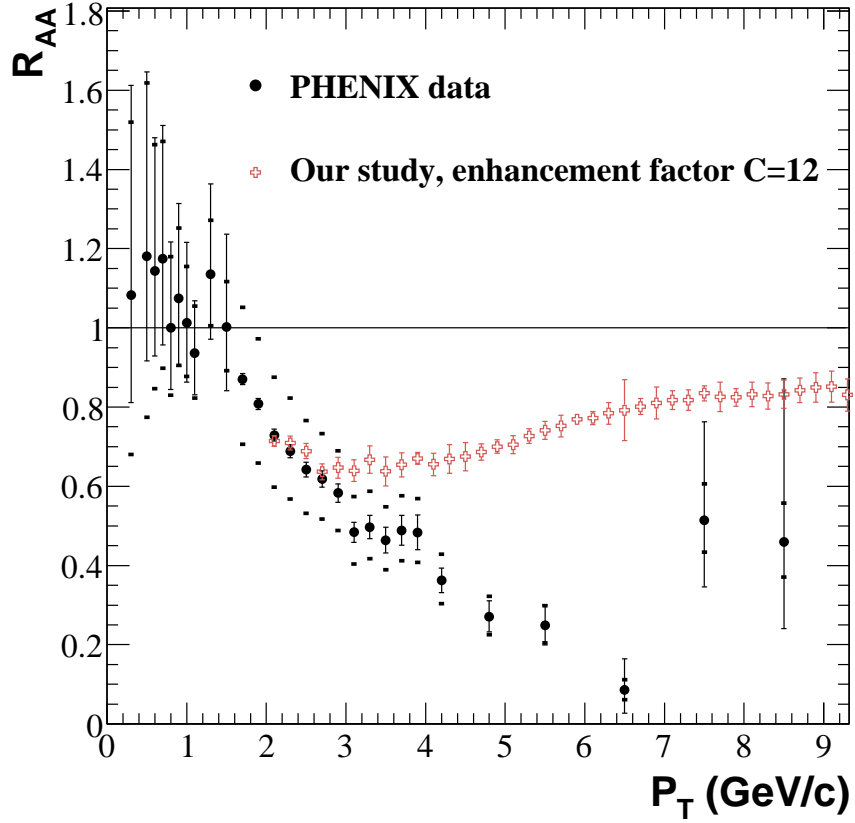,width=0.75\linewidth}
 \caption{Nuclear modification factor of non-photonic electrons (dots) 
	measured in central ($0-10\%$) Au+Au collisions at 
	$\sqrt{s_{NN}} = 200~{\rm GeV}$~\protect\cite{Adare:2006nq}.
	The crosses correspond to the results of the 
	simulation described in the text for a $\Lambda_c/D$ 
	enhancement factor of 12.}
 \label{figure3} 
\end{figure}

As stated above, the non-photonic electron $p_{\rm t}$ distribution in
Au+Au collisions has been evaluated 
after considering a enhancement whose shape is, according to 
the predictions of the coalescence model~\cite{Greco:INFN2005}, assumed to be 
a Gaussian versus $p_{\rm t}$.
It has the following parameters. 
Mean: $5~{\rm GeV}/c$, constant: $\sim 0.9$ and sigma: $2.9~{\rm GeV/c}$.
The constant of 0.9 is obtained from 
$N_{\Lambda_c,\bar{\Lambda}_{c}}/N_D \times {\cal C}$ with 
$N_{\Lambda_c,\bar{\Lambda}_{c}}/N_D = 7.3\%$ (Tab.~\ref{tab_PDG}) 
and ${\cal C} = 12$.
Such an enhancement factor ${\cal C} = 12$ is justified since the 
resulting $\Lambda_{c}/D$ ratio of $\sim 0.9$ is of the same order of magnitude
as the non-strange and strange baryon/meson ratios 
measured by the \textsc{STAR} collaboration~\cite{Sarah}.
In contrast, the corresponding (enhanced) $\Lambda_{c}/D$ 
ratio is in ref.~\cite{Sorensen:2005sm} located at lower $p_{\rm t}$ and 
its maximum is close to 1.7.
The enhancement is applied such that the $p_{\rm t}$-differential charm 
cross-section is conserved.
The latter is an arbitrary choice that could be justified since most of the 
charm hadron transverse momentum is given by the charm quark whatever, 
baryon or meson, this hadron is.
We finally compute the $R_{AA}$ ratio from the non-photonic electron
$p_{\rm t}$ spectra 
assuming that the only medium induced effect is the $\Lambda_c/D$ enhancement. 
The results are shown in Fig.~\ref{figure3} together with the 
\textsc{PHENIX} data.
Note that at this step only electrons from charm decay are considered 
and heavy-quark energy loss is neglected.
The simulated $R_{AA}$ ratio is shown only for $p_{\rm t} > 2~{\rm GeV/c}$ 
since shadowing might play a role and has not been considered in the 
simulation.
One can see that the $\Lambda_c/D$ ratio close to unity in central collisions 
at $p_{\rm t} = 5~{\rm GeV/c}$ can already explain 
$\sim 40\%$ of the suppression of non-photonic electrons in the 
$2-4~{\rm GeV}/c$ $p_{\rm t}$ range.
Even in the high $p_{\rm t}$ region ($8-9~{\rm GeV}/c$) the $\Lambda_c/D$
enhancement results in a significant suppression of non-photonic electrons.

\begin{figure}[htb]
  \centering\epsfig{file=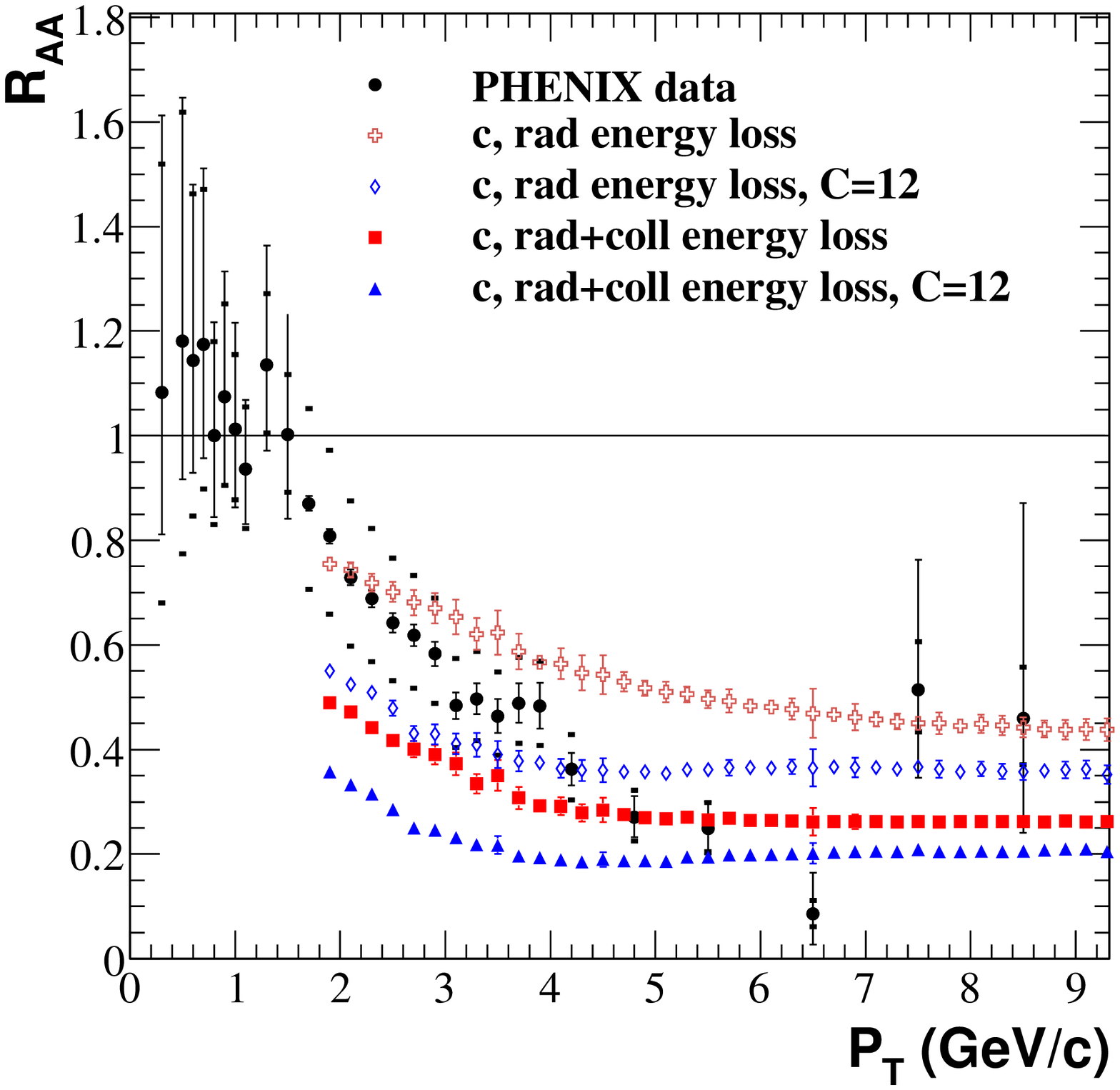,width=0.75\linewidth}
 \caption{Nuclear modification factor of non-photonic electrons (dots) 
	measured in central ($0-10\%$) Au+Au collisions at 
	$\sqrt{s_{NN}} = 200~{\rm GeV}$~\protect\cite{Adare:2006nq}.
	The symbols show the result of the simulation described in the text.}
 \label{figure4} 
\end{figure}

In the next step $c$ quark radiative and collisional energy loss is included 
in the simulation.
This is achieved by a convolution of our non-photonic electron $p_{\rm t}$ 
spectra with the differential suppression factors taken 
from~\cite{Wicks:2005gt}. 
It is shown in Fig.~\ref{figure4} that the relative suppression 
originating from the $\Lambda_{c}/D$ enhancement is about the same amplitude 
than the one from the charm collisional energy loss and represents about 
$36(20)\%$ of the observed suppression at $p_{\rm t} = 4(9)~{\rm GeV}/c$.
In contrast, the suppression reported in~\cite{Sorensen:2005sm} is 
less than $20\%$ in the $p_{\rm t}$ range $2-5~{\rm GeV}/c$ and 
becomes negligible for $p_{\rm t} > 5~{\rm GeV}/c$.

\begin{figure}[htb]
  \centering\epsfig{file=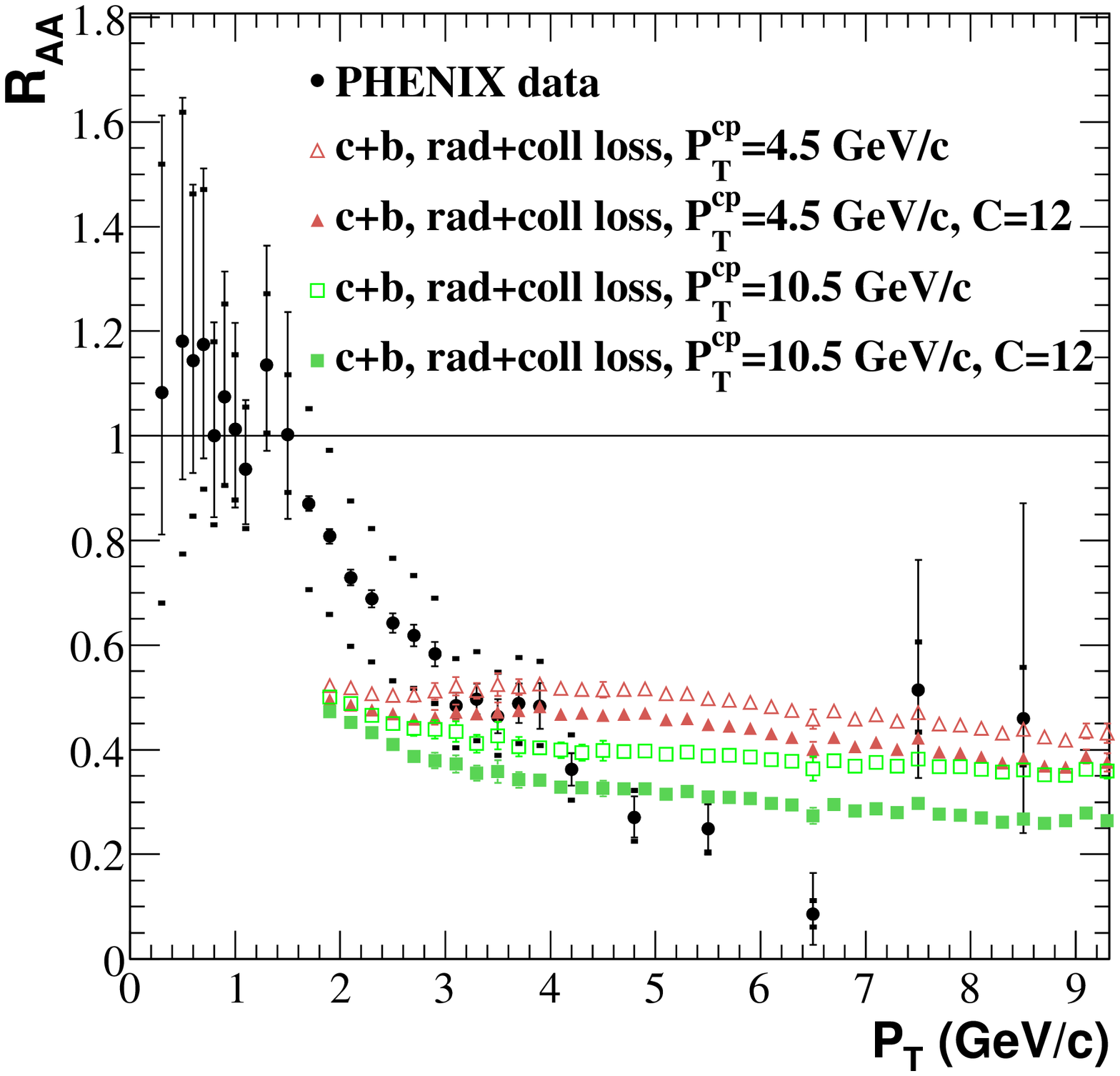,width=0.75\linewidth}
  \caption{Nuclear modification factor of non-photonic electrons (dots) 
	measured in central ($0-10\%$) Au+Au collisions at 
	$\sqrt{s_{NN}} = 200~{\rm GeV}$~\protect\cite{Adare:2006nq}.
	The symbols show the result of the simulation described in the text.}
 \label{figure5} 
\end{figure}

Finally the bottom contribution is added in the simulation.
This obviously reduces the suppression of the sum of non-photonic electrons
because $b$ quarks are supposed to lose less energy than $c$ quarks.
However, the relative contribution of $c$ and $b$ quarks to the total 
non-photonic electron yield is not well known.
According to FONLL predictions, the crossing point ($p_{\rm t}^{cp}$) 
between charm and bottom electron decay $p_{\rm t}$ spectra is expected 
to be located in the range $2.5 < p_{\rm t} < 10.5~{\rm GeV}/c$.
Therefore we have considered two scenarios to include the bottom contribution:
a crossing point in the central value predicted by FONLL 
($p_{\rm t}^{cp} = 4.5~{\rm GeV}/c$) and the highest possible crossing 
point allowed by the calculation ($p_{\rm t}^{cp} = 10.5~{\rm GeV}/c$).
The latter results in the weakest contribution of electrons from $b$ quark 
decay to the total non-photonic electron yield.
As shown in Fig.~\ref{figure5}, 
whatever the assumed crossing point, the effect of the $\Lambda_c/D$ 
enhancement remains visible.
It leads to a decrease of the non-photonic electron $R_{AA}$ of about 
10(25)\% for a crossing point at $p_{\rm t}^{cp} = 4.5(10.5)~{\rm GeV}/c$.

In addition to the $\Lambda_c/D$ enhancement addressed in this work, 
one could expect an enhancement of the $D_s/D$ ratio 
due to the strangeness enhancement in heavy ion collisions. 
According to Tab.~\ref{tab_PDG}, the $D^{\pm}_{s}$ semi-leptonic decay 
branching ratio is similar to that of $D^{0}$ which represents the main source 
of non-photonic electrons. 
An enhancement of $D_{s}^{\pm}$ mesons would 
therefore not affect significantly the $R_{AA}$ of non-photonic electrons. 
However, as the uncertainty on the measured semi-leptonic decay branching ratio
of $D^{\pm}_{s}$ mesons is large (Tab.~\ref{tab_PDG}), the contribution
of $D^{\pm}_{s}$ mesons to the non-photonic electron yield and consequently
to the non-photonic electron $R_{AA}$ cannot be estimated precisely.
From the theoretical side, according to the Spectator Model 
for charm mesons decay, the semi-leptonic decay widths for the different 
charm mesons should be equivalent~\cite{Altarelli:1982kh}. 
Knowing charm meson lifetimes~\cite{Yao:2006px} 
and branching ratios for $D^0$ and $D^\pm$ (see Tab.\ref{tab_PDG}), 
one can estimate the $D^\pm_s$ branching ratio to 
$8.2 \pm 0.2\%$ which appears to be consistent with the measured value.

In summary, we have shown that an enhancement of the $\Lambda_c/D$ ratio 
in heavy ion collisions has important consequences on the nuclear modification 
factor of non-photonic electrons.
Such an enhancement, which has recently been predicted by the coalescence model
and which has already been measured for non-strange and 
strange hadrons, would significantly reduce the $R_{AA}$ of non-photonic 
electrons at intermediate $p_{\rm t}$.
This is a consequence of the smaller semi-leptonic decay branching ratio
of charm baryons compared to that of charm mesons
and of the softer decay lepton spectrum from charm baryons compared to that 
of charm mesons.
In the most realistic situation investigated in the present work the
enhancement leads to an additional non-photonic electron suppression of 
$10-25\%$ (with respect to the suppression observed without charm 
baryon/meson enhancement). 
This suppression can even be larger in case of a weaker bottom contribution 
to the non-photonic electron spectrum.
We conclude that it is therefore premature to interpret the non-photonic 
electron $R_{AA}$ data before a possible enhancement of the $\Lambda_c/D$ ratio
is measured experimentally.
Heavy quark energy loss can be studied in a much cleaner way via the nuclear 
modification factor of exclusively reconstructed charm hadrons.
Such measurements should be possible with the RHIC-II
experiments~\cite{jacak} and with the ALICE experiment at the 
LHC~\cite{Alessandro:2006yt}.
We finally note that the $\Lambda_c/D$ enhancement can possibly influence
the elliptic flow of non-photonic electrons as well.

\acknowledgements
We gratefully acknowledge Anton Andronic, Pol-Bernard Gossiaux and 
Peter Levai for carefully reading the manuscript and for valuable suggestions 
and St\'ephane Peign\'e and Vincenzo Greco for fruitful discussions.

\begin{center}
\begin{table}[h]
\caption{Inclusive decay branching ratio ($BR$) of charm hadrons
into $e$ $+$ anything~\protect\cite{Yao:2006px} and yield ($N$)
of charm hadrons (in 4$\pi$) in $pp$ collisions at $\sqrt{s} = 200~{\rm GeV}$ 
from the present \textsc{PYTHIA} simulation using input parameters as described in~\protect\cite{Adcox:2002cg}.
The total cross-section for charm production is normalized to the 
experimental value obtained in~\protect\cite{Adcox:2002cg}.
$N_{\Lambda_c}$ and $N_{\bar{\Lambda}_c}$ include primarily produced 
$\Lambda_c$ and $\bar{\Lambda}_{c}$ as well as those from $\Sigma_c$ and 
$\bar{\Sigma}_{c}$ decay. $\epsilon_{N_e} (BR)$ is the contribution to the
uncertainty of the total electron yield due to the uncertainty on the 
particle $BR$.}
\begin{tabular}{lllllllll}
Hadron        & $D^+$ & $D^-$ & $D^0$ & $\bar{D^0}$ & $D^+_s$ & $D^-_s$ &
$\Lambda_c$ & $\bar{\Lambda_c}$ \\
\hline
$BR$ (\%) & \multicolumn{2}{c}{$17.2\pm 1.9$} & 
	  \multicolumn{2}{c}{$6.71\pm 0.29$} & 
          \multicolumn{2}{c}{$8^{+6}_{-5}$} & 
          \multicolumn{2}{c}{$4.5\pm 1.7$}\\ 
\hline 
$N$ $(\times 10^{-3}$) & 3.00 & 3.07 & 9.31 & 9.85 & 1.82 & 1.60 & 1.23 & 0.85 \\
\hline
$\epsilon_{N_e} (BR)$ (\%) & 1.08 & 1.10 & 1.31 & 1.39 & 4.41 & 3.58 & 1.51 & 1.04 \\
\end{tabular}
\label{tab_PDG}
\end{table}
\end{center}

\newpage

\begin{center}
\begin{table}[h]
\caption{Main differences between the approach presented
in~\protect\cite{Sorensen:2005sm} and this work. See text for more details.}
\begin{tabular}{lll}
& \protect\cite{Sorensen:2005sm} & this work \\
\hline
$\Lambda_c/D$ versus $p_{\rm t}$ in Au+Au collisions & as $\Lambda/K^0_s$ data & Gaussian \\
\hline
$\Lambda_c/D$ versus $p_{\rm t}$ in $pp$ collisions & as $\Lambda/K^0_s$ data & \textsc{PYTHIA} \\
\hline
Maximum of the $\Lambda_c/D$ enhancement & $\sim 1.7$ at $p_{\rm t} \sim 3~{\rm GeV}/c$  
& $\sim 0.9$ at $p_{\rm t} \sim 5~{\rm GeV}/c$\\
\hline
Energy loss &  hadron shape scaling & \protect\cite{Wicks:2005gt} \\
\hline
Electrons from bottom decay &  no & yes \\
\end{tabular}
\label{tab_diff}
\end{table}
\end{center} 

\newpage

\newpage

\newpage

\newpage

\newpage

\end{document}